\documentclass[%
reprint,
 amsmath,amssymb,
 aps,10pt,
showkeys,
]{revtex4-1}

\usepackage{graphicx}
\usepackage{dcolumn}
\usepackage{bm}
\usepackage{float}
\usepackage{color}
\usepackage{hyperref}
\usepackage{soul}
\usepackage{enumitem}
\hypersetup{colorlinks=true, citecolor=blue, urlcolor=blue, linkcolor=blue}
\usepackage{placeins}

\graphicspath{{figures/}}

\begin{document}

\preprint{APS/123-QED}

\title{Enhancing social cohesion with cooperative bots in societies of greedy, mobile individuals}


\author{Lei Shi$^{1,2}$}
\thanks{These authors contributed equally.}
\author{Zhixue He$^{1,4}$}
\thanks{These authors contributed equally.}
\author{Chen Shen$^3$}
\email{steven\_shen91@hotmail.com}
\author{Jun Tanimoto$^{3,4}$}
\affiliation{
1. School of Statistics and Mathematics, Yunnan University of Finance and Economics, 650221, Kunming, China.\\
2. Interdisciplinary Research Institute of data science, Shanghai Lixin University of Accounting and Finance, 201209, Shanghai, China.\\
3. Faculty of Engineering Sciences, Kyushu University, Fukuoka, 816-8580, Japan.\\
4. Interdisciplinary Graduate School of Engineering Sciences, Kyushu University, Fukuoka, 816-8580, Japan.
}

\date{\today}

\begin{abstract}
Addressing collective issues in social development requires a high level of social cohesion, characterized by cooperation and close social connections. However, social cohesion is challenged by selfish, greedy individuals. With the advancement of artificial intelligence (AI), the dynamics of human-machine hybrid interactions introduce new complexities in fostering social cohesion. This study explores the impact of simple bots on social cohesion from the perspective of human-machine hybrid populations within network. By investigating collective self-organizing movement during migration, results indicate that cooperative bots can promote cooperation, facilitate individual aggregation, and thereby enhance social cohesion. The random exploration movement of bots can break the frozen state of greedy population, help to separate defectors in cooperative clusters, and promote the establishment of cooperative clusters. However, the presence of defective bots can weaken social cohesion, underscoring the importance of carefully designing bot behavior. Our research reveals the potential of bots in guiding social self-organization and provides insights for enhancing social cohesion in the era of human-machine interaction within social networks.
\end{abstract}


\maketitle


\section{Introduction}

Social cohesion represents a crucial collective consciousness in contemporary societies facing collective issues such as epidemics \cite{ash2022disease}, economic crises \cite{acharya2009causes}, social inequality \cite{hauser2019social}, and climate change \cite{steffen2015planetary}. However, the establishment of a highly cohesive social system is often hindered by individual selfish behaviors \cite{tainter1988collapse,schiefer2017essentials}. So far, some studies employing evolutionary game theory \cite{sachs2004evolution,nowak2006evolutionary} have revealed the generation and maintenance of social cohesion through self-organizing processes in interpersonal interactions within mobile population \cite{schiefer2017essentials,helbing2011self,castellano2009statistical,fu2013global,wu2012expectation,wu2011moving}, aiming to explore ways to enhance social cohesion \cite{roca2011emergence}. 
With the advent and integration of artificial intelligence (AI) technology in social settings, AI-driven agents, or bots, are becoming a part of social fabric \cite{abdullah2022chatgpt,santos2024prosocial,chen2023ensuring}, shifting traditional human-to-human interactions to a new paradigm of human-machine hybrid interactions \cite{crandall2018cooperating,shirado2020interdisciplinary,santos2019evolution,shirado2017locally,shirado2020network,mckee2023scaffolding,hilbe2014extortion}. The impact of this shift in interaction on social cohesion and collective behavior remains unclear. This study aims to deepen the understanding of the influence of bots on social cohesion by investigating self-organized movements within human-machine hybrid populations, and to explore potential avenues for enhancing social cohesion within such hybrid population contexts.

Social cohesion is fundamentally composed of two elements: \textit{orientation towards the common goods} and \textit{nurturing of social relationships} \cite{schiefer2017essentials}. The \textit{orientation towards the common goods} involves cooperative behavior that individuals prioritizing collective benefits over personal gains \cite{sachs2004evolution,nowak2011supercooperators}. It is often challenged by selfish free-riding behavior, leading to the ``tragedy of the commons'' \cite{hardin1968tragedy}. Over the last few decades, various reciprocity mechanisms that support evolution of cooperation have been revealed. These include direct reciprocity, which arises from repeated interactions~\cite{schmid2021unified}, indirect reciprocity that replies on the behavioral information transmission \cite{rockenbach2006efficient}, and network reciprocity, which is influenced by the structure of interactions~\cite{nowak2006five}.

Migration is a fundamental behavior among individuals, providing a perspective for exploring the maintenance and the formation of \textit{social relationships}.  
Traditionally, migration refers to physical movement, such as residential relocation. However, in the information technology era, research now includes digital network migration behavior\cite{kumar2011understanding,newell2016user,fiesler2020moving}. In online networks like \textit{Reddit} and \textit{GitHub}, users can freely switch between working groups and communities, altering their interactional relationships as desired. These migration promotes the formation of self-organized movements\cite{vainstein2007does,helbing2008migration}. Different social interaction, including segregation \cite{hamnett2001social}, interweaving \cite{he2022q}, and aggregation \cite{helbing2009outbreak,roca2011emergence}, arise in self-organizing movements driven by various migration preferences. In particular, individuals driven by payoffs tend to form close social bonds through movement \cite{roca2011emergence}. This spatial clustering during migration can enhance cooperation, but depends on specific conditions, such as moderate population density or low mobility \cite{vainstein2007does,helbing2009outbreak}. Conversely, excessive greed for personal interests impedes the development of cooperation, disrupts the establishment of social connections, and ultimately erodes social cohesion \cite{roca2011emergence}.

\begin{figure*}[ht]
    \centering
    \includegraphics[width=0.7\linewidth]{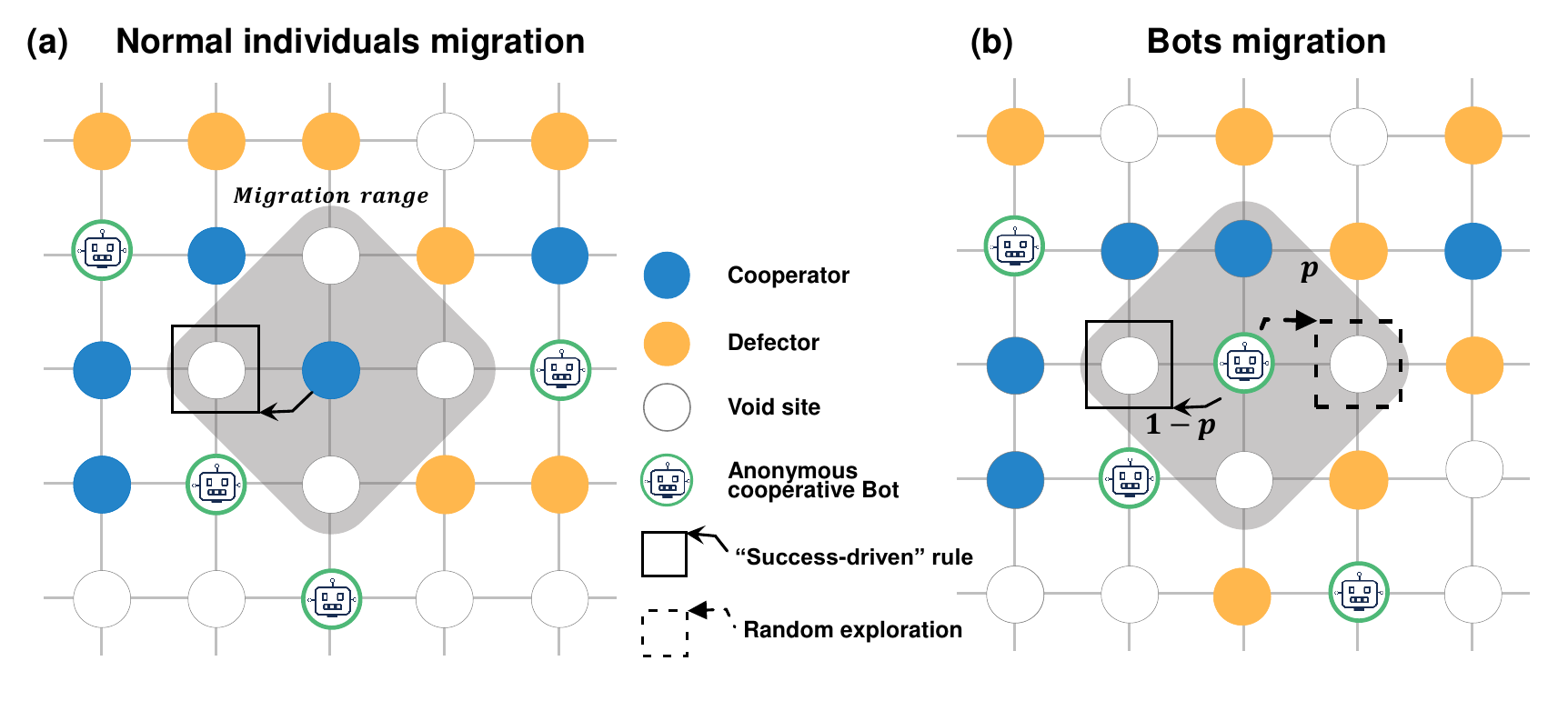}
    \caption{ \textbf{Schematic diagram of model setting.} A hybrid population of size $n$, consisting of proportional $\phi$ mobile bots and $1 - \phi$ normal individuals deployed on a lattice network of size $N$ where some nodes on the network are empty ($n < N$) and available for agents to migrate. Normal individuals mimic the strategy of most successful neighbors (including themselves) and migrate based on the ``success-driven'' rule, moving to adjacent unoccupied location (within the shaded area) that generate the highest payoffs in fictitious play. Anonymous Bots are programmed to engage in unconditional cooperation, and migrate randomly with a probability of $p$, while follow ``success-driven'' rule with a probability of $1-p$.}
    \label{fig1_model}
\end{figure*}

Recent research has sparked interest in using AI-driven agents or bots to study cooperation issues \cite{guo2023facilitating,Sharma2023small}. They have revealed bots' ability to address coordination dilemmas \cite{shirado2017locally} and scaffold cooperation \cite{mckee2023scaffolding} by integrating bots into network engineering and game interactions \cite{shirado2020network,hilbe2014extortion}. Here, our focus extends beyond the influence of bots on individual cooperation to their role in the collective self-organization movements within populations. To achieve this, we introduce bots into a mobile populations comprised of selfish, greedy individuals whose behavioral decisions aim to maximize personal gains. Our model does not assume bots possess complete knowledge of individual behavior or engage in coordinated actions towards normal individuals \cite{shirado2017locally,shirado2020network}. Instead, we enable bots to participate autonomously in migration, using a simple behavioral design characterized by consistent adoption of cooperative action and mobile exploration with some randomness. As we will see, these cooperative bots can facilitate cooperation and fostering spatial clustering within mobile populations, thereby promoting highly social cohesion. Interestingly, cooperative bots can  break the population out of its frozen state, stimulating the self-organization movement among selfish, greedy individuals. The emergence and maintenance of social cohesion in sparse mobile populations have often been a challenge in human-human interactions, but the introduction of cooperative bots can solve this challenge situation in such hybrid populations. Therefore, our study reveal the potential of simple cooperative bots in guiding individual behavior to address collective issues.

\section{Model}

\paragraph*{\textbf{Hybrid population}} 
We investigate a hybrid population consisting of both bots and greedy normal players, with proportions $\phi \in [0, 0.5]$ and $1-\phi$ respectively. This population is placed on a grid lattice network of size $N=L\times L$ with periodic boundary and $K$-nearest neighboring sites (specifically focusing on the von Neumann neighborhood where $K=4$). Each site in the network can be occupied by either a bot/normal player, or it may remain unoccupied, so we define the population density as $\rho=n/N$ ($0<\rho <1$). Consequently, the network contains $N \rho \phi$ bots and $N \rho (1-\phi)$ normal players.

Our model is conducted using Monte Carlo (MC) asynchronous simulations. In each MC time step, both of bots and normal players undergo three stages : game interaction, strategic updating, and migration. The model architecture is depicted in Fig.~\ref{fig1_model}.

\paragraph*{\textbf{Game interaction}} 
Game interactions are implemented through a one-shot PD game. Both of bots and normal players participate in the paired game with their neighbors, making decisions to adopt either unconditional cooperation strategy ($C$) or unconditional defection strategy ($D$). Cooperation means incurring a cost $c$ to benefit others by an amount $b$, while defection does nothing. Mutual cooperation yields a reward of $R=b-c$, while mutual defection leads to a punishment of $P=0$ for both agents. A cooperator receives a sucker's payoff of $S=-c$, whereas a defector gains a temptation to defect payoff of $T=b$ upon meeting each other. To simplify the model without loss generality, we define the dilemma strength as $r=c/(b-c)$ and set $b-c=1$ following the method outlined in ref. \cite{wang2015universal}. The payoff matrix is then re-scaled as:
\begin{equation}
    \begin{array}{lc}
    \mbox{}&
    \begin{array}{cc}C\ \ \ \ &\ \ D \ \  \end{array}\\
    \begin{array}{c}C\\ D  \end{array}&
    \left(\begin{array}{cc}
    1 & -r  \\
    1+r & 0  \\
    \end{array}\right).
    \end{array}
\end{equation}
We focus on the influence of cooperative bots which are programmed to consistently choose unconditional cooperation without altering their behavior in interaction. 
We also explore defective bots which consistently choose unconditional defection, and the corresponding outcomes are presented in the ``Supplementary Information'' (SI).

\paragraph*{\textbf{Strategic updating}} 
We utilize an anonymous setup where normal players remain unaware of the presence of bots, thereby exclude potential bias against the bots among them \cite{karpus2021algorithm,ishowo2019behavioural,wang2017onymity}. The decision-making of greedy normal players is driven by the maximization of their own profits. They employ the "best-take-over" rule \cite{nowak2011supercooperators}, imitating the strategy of their neighbors that yields the highest payoff when their neighbors' payoff exceeds their own.

\paragraph*{\textbf{Migration}} 

Normal players strategically also migrate to maximize their payoffs by adhering to the ``success-driven'' rule \cite{helbing2008migration, helbing2009outbreak}. They move to adjacent vacant sites (including their current position) that offer the highest payoffs in fictitious play. To prevent bots from being consistently exploited by their counterparts and to enhance adaptation to the migration environment, bots are programmed to engage in exploratory migration. Bots take random movements with a probability of $p \in[0,1]$ and follow the ``success-driven''  rule with a probability of $1-p$.

To analyze the impact of bots on social cohesion, we utilize the fraction of cooperation among normal players ($F_C$) as as a metric to evaluate how bots promote individuals to consider collective interests. 
On the other hand, we assess the overall spatial aggregation of normal players, denoted as $Agg$, to gain insights into the influence of bots on social cohesion from a spatial perspective. $Agg$ is the weighted average of the aggregation levels of normal cooperators ($Agg_{C}$) and defectors ($Agg_{D}$), expressed as:
\begin{gather}
    Agg_{i} = \frac{L_{ i \mbox{-} C } + L_{ i \mbox{-} D }}{K}\ \ \ \  i\in\{C,D\}, \\
    Agg = F_C \times Agg_{C} + (1-F_C) \times Agg_{D},
\end{gather}
where $L_{ x \mbox{-} C }$ ($L_{ x \mbox{-} D }$) is the average number of neighboring cooperators (defectors) for normal players who adopt strategy $x$.

To investigate how bots influence the self-organization dynamics during migration, inspired by the cluster shape analysis detailed in ref.~\cite{fu2010invasion}, we introduce $\lambda_C \in [-1, 1]$ and $\lambda_D \in [-1, 1]$ as indices to assess the clustering level of normal cooperators and the separation level of normal defectors relative to the cooperative clusters, respectively: 
\begin{gather}
    \lambda_C = \frac{1}{|\Omega_C|} \sum_{i \in \Omega_C } \frac{m^{i}_C - m^{i}_D}{K} ,\\
    \lambda_D = \frac{1}{|\Omega_D|} \sum_{i \in \Omega_D } \frac{m^{i}_{void} - m^{i}_C}{K},
\end{gather}
where $\Omega_C$ and $\Omega_D$ denote the sets of normal cooperators and normal defectors, respectively, with their respective quantities denoted by $|\Omega_C|$ and $|\Omega_D|$. $m^{i}_{C}$, $m^{i}_{D}$, and $m^{i}_{void}$ are the numbers of cooperators, defectors, and void sites within the neighborhood of player $i$, respectively. A $\lambda_C$ approaching 1 suggests cooperators can form tightly clusters. Conversely, a greater number of links between cooperators and defectors result in $\lambda_C < 0$, indicating a negative assortment among cooperators \cite{fu2010invasion}. For $\lambda_D>0$, defectors are adjacent to cooperative clusters, and as $\lambda_D$ approaches 1, it indicates that the degree of separation between defectors and cooperative clusters increases. Conversely, a negative $\lambda_D$ implies defectors are embedded within the cooperative clusters.

\begin{figure}[b]
    \centering
    \includegraphics[width=0.9\linewidth]{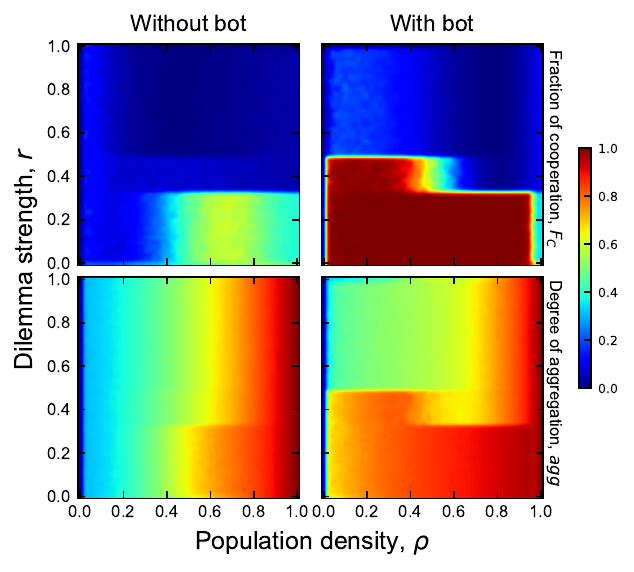}
    \caption{ \textbf{Simple cooperative bots facilitate cooperation and population aggregation, promoting the establishment of highly cohesive collective behavior.} The color code indicates the fraction of cooperation (top panels) and the degree of aggregation of normal individuals (bottom panels) as a function of population density $\rho$ and temptation $r$. Parameters are set to $\phi=0.5$ and $p=0.01$ for the scenario with bots.}
    \label{fig1_r_rho}
\end{figure}

For our computer simulation, we maintained a fixed grid size of $N = 100 \times 100$. To ensure the reliability and stability of our results, we averaged the final outcomes over 50 independent runs. Each run involved averaging the last $5000$ time steps out of more than $10^6$ Monte Carlo (MC) time steps. To confirm the robustness of our model and the obtained results, we extensively explored various scenarios, detailed in SI. This exploration encompassed different proportions of bots, varied migration and interaction strategies for bots, limited mobility of normal players, and the effects of lattice network size and its topological structure. We investigate the influence of bots across different levels of complete information acquisition, taking into account the strategic decisions of normal players. Furthermore, we examine the scenarios where the decision-making processes of normal individuals involve behavioral noise within the SI, aiming to validate the robustness of the results obtained by relaxing the assumption of absolute rationality among normal individuals presented in the main text.

\begin{figure*}[ht]
    \centering
    \includegraphics[width=0.74\linewidth]{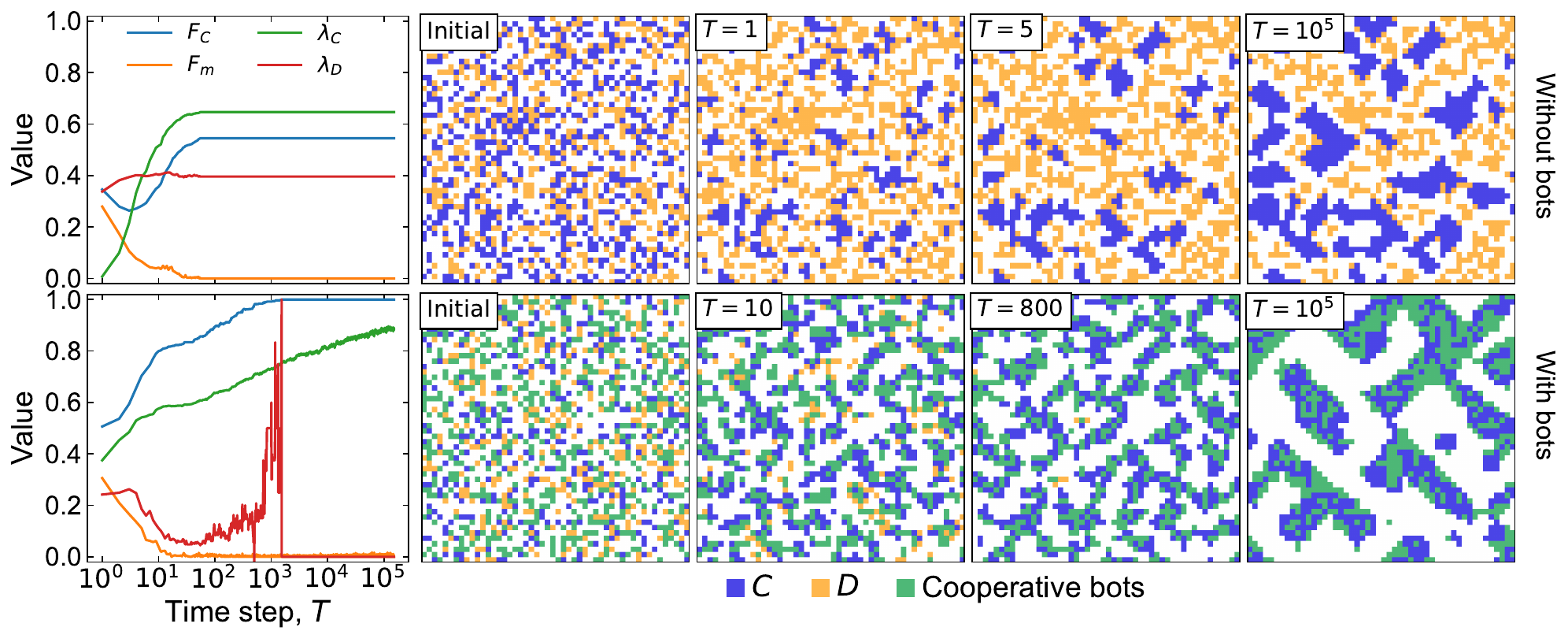}
    \caption{\textbf{ 
    Cooperative bots can drive self-organized movement of normal individuals, preventing the population from entering a frozen state that would typically occur in their absence.} Leftmost panels show the fraction of cooperation $F_C$, the fraction of normal individuals who have moved $F_m$, clustering level of normal cooperators $\lambda_C$ and the isolation level of normal defectors $\lambda_D$ as a function of time steps $T$. The right panels showcase evolutionary snapshots in scenarios with and without bot, respectively. In these snapshots, blue, yellow, green, and white dots represent cooperators, defectors, bots, and empty sites, respectively. Results are obtained by setting $\rho = 0.5$, $r=0.2$ and $p=0.01$.}
    \label{fig2_evo1}
\end{figure*}

\section{Results}

\subsection{Simple bot promote social cohesion}

We begin by examining the influence of cooperative bots. Previous studies have shown that migration can facilitate the clustering of cooperators in populations of moderate density, thereby enhancing network reciprocity when the dilemma strength is low (i.e., $r < 1/3$) \cite{roca2011emergence,meloni2009effects}, as depicted in the left panels of Fig.~\ref{fig1_r_rho}. However, in sparse populations (i.e., $\rho < 0.4$), the abundance of empty sites hampers the formation of cooperative clusters, resulting in the decline of cooperation. Under high levels of dilemma strength (i.e., $r > 1/3$), cooperation cannot be sustained even with available empty nodes and individual migration. Interestingly, the introduction of cooperative bots significantly promote cooperation among normal individuals, see the right panels of Fig.~\ref{fig1_r_rho}. Compared to scenarios without bots, the presence of bots greatly enhances cooperation across a wider range of population density parameters when $r < 1/3$, ensuring high levels of cooperation in both sparse and dense populations. Even under high dilemma strength, bots prove effective in maintaining cooperation. Furthermore, only a few bots are needed to have a significant impact on individual behavior, see Fig.~\ref{figs_phi} in the SI. At a moderate population density of $\rho = 0.6$, introducing a minority of mobile bots is sufficient to sustain a high level of cooperation (i.e., approximately $\phi \approx 0.02$ for random mobile bots and $\phi \approx 0.11$ for low-exploration bots with $p=0.01$). Even under high dilemma strength, as low as $\phi \approx 0.2$ proportion of low-exploration bots can maintain cooperation.

\begin{figure}[t]
    \centering
    \includegraphics[width=0.9\linewidth]{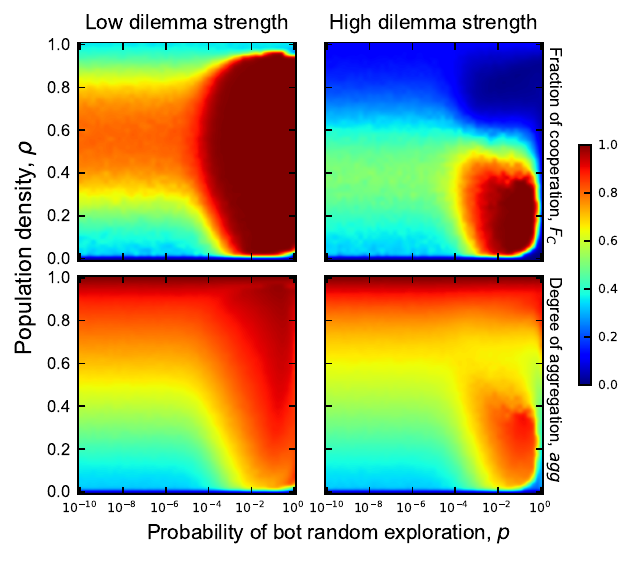}
    \caption{\textbf{Bots with a moderate level of random exploration can contribute to the promotion of social cohesion.} The color code indicates the fraction of cooperation (top panels) and the degree of aggregation of normal individuals (bottom panels) as a function of population density $\rho$ and the probability of bot random exploration $p$ under low temptation $r=0.2$ and high temptation $r=0.4$. The fraction of bot is set to $\phi=0.5$.}
    \label{fig3_prob}
\end{figure}

In sparse populations, the abundance of empty nodes separating individuals leads to high isolation and low aggregation. As population density increases, a corresponding increase in the degree of aggregation as expected. Bots also can promote population clustering. Even at low population densities (i.e., $\rho < 0.4$), bots can induce normal individuals to aggregate, resulting in a high degree of aggregation, as shown in the bottom panels of Fig.~\ref{fig1_r_rho}. Similarly, a minority of bots can significantly enhance the degree of aggregation, as depicted in Fig.~\ref{figs_phi_agg} in the SI. These findings demonstrate that cooperative bots not only facilitate cooperation but also aggregate individuals, thereby promoting the emergence of high levels of social cohesion.

\begin{figure*}[th]
    \centering
    \includegraphics[width=0.74\linewidth]{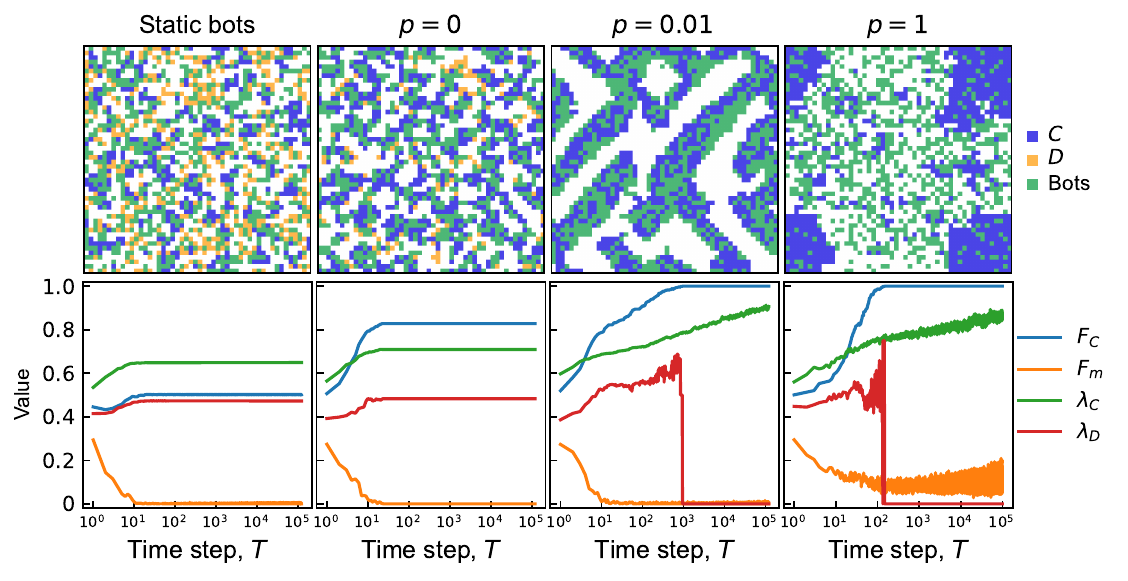}
    \caption{\textbf{ The random exploration of mobile bots drives the separation of defectors from cooperative clusters and facilitates the aggregation of cooperators.} Top panels show stable spatial distribution in scenarios with static bots (i.e., unable to migrate), success-driven bots (i.e., $p=0$), low-exploration bots (i.e., $p=0.01$) and randomly migrating bots (i.e., 
 $p=1$). Bottom panels show that the fraction of cooperation $F_C$, the fraction of normal individuals who have moved $F_m$, clustering level of normal cooperators $\lambda_C$ and the isolation level of normal defectors $\lambda_D$ as a function of time steps $T$, respectively. All stable spatial distribution are obtained at $T=10^5$. These outcomes were obtained with parameter settings of $\rho = 0.6$, $r=0.2$ and $\phi=0.5$. }
    \label{fig4_evo2}
\end{figure*}

\subsection{Self-organizing movement}

To understand the influence of mobile bots on social cohesion, Fig.~\ref{fig2_evo1} depicts the temporal evolution and spatial distribution of the population. The dynamic visualization of this process is available online at \href{https://osf.io/tu6cq}{https://osf.io/tu6cq}. In the absence of bots, a typical evolutionary process unfolds (shown in the top panels of Fig.~\ref{fig2_evo1}). Initially, the random spatial distribution impedes the survival of isolated cooperators, leading to a decline in the fraction of  cooperation ($F_C$) in the early stages of evolution. When migration is feasible, cooperative migration further drives the aggregation into clusters, compared to the cluster formation process without migration as discussed in \cite{fu2010invasion}. While empty nodes may partition some defectors and cooperators, they also constrain further expansion of cooperators. When no more profitable positions exist, normal individuals cease movement (with the fraction of normal individuals who moved $F_m$ remaining at zero), leading to a frozen state in  self-organizing movement of greedy population\cite{helbing2009outbreak}.

Interestingly, the introduction of cooperative bots disrupts this frozen state (although the value of $F_m$ is low, it is not zero), as depicted in the bottom panels of Fig.~\ref{fig2_evo1}. It indicates that the presence of bots indirectly creates profitable positions, driving normal individuals to move, and thus break the frozen state. More importantly, cooperative bots can facilitate tight cooperator cluster formation (evidenced by continuous $\lambda_C$ increase) while driving defector separation from cooperative clusters (as shown by rising $\lambda_D$). This self-organizing movement, catalyzed by cooperative bots, renders defectors defeated by tightly-knit cooperative clusters. Even after defectors vanish, bots can further facilitate the aggregation of cooperator (The final value of $\lambda_C$ stabilizes at a high level). In the presence of behavioral noise, individuals randomly reset their strategies and migration with a certain probability. These stochastic behaviors can prevent the system from reaching a complete freeze \cite{helbing2009outbreak}. Intriguingly, cooperative bots also can foster the levels of cooperation and aggregation among the population, compare to the scenarios in absence of bots, as illustrated in Figs.~\ref{figs_nois} and \ref{figs_evo} in the SI.

However, in extremely dense populations (i.e., $\rho=0.97$), cooperative bots fail to eliminate defectors as they cannot efficiently drive defector separation from cooperative clusters, see top panels of Fig.~\ref{fig_evo_condi} in SI. On the other hand, in extremely sparse populations (i.e., $\rho<0.02$), cooperative clusters cannot be established. Under a high dilemma strength (i.e., $r=0.4$), the self-organization movement promoted by the bot cannot eliminate defection due to the payoffs advantage of the defector, see bottom panels of Fig.~\ref{fig_evo_condi} in SI.  It is worth noting that the normal individuals' strategy updating and ``success-driven'' migration depends on access to complete information about others' behaviors. When avenues for acquiring such comprehensive information are restricted, individuals rely solely on their own judgment for strategy updating (employing the myopic principle \cite{nowak2011supercooperators}, wherein normal individuals tend to adopt a better response strategy in the current situation) and resort to random migration. Results show that cooperative bots still can to promote cooperation under conditions of limited information, as depicted in Fig.~\ref{figs_infor} of SI. Nevertheless, cooperative bots no longer effectively aggregate normal individuals, as illustrated in Fig~\ref{figs_fb} of SI.

\subsection{The effects of bot behavior}

The mobility of bots is instrumental in driving the self-organizing dynamics of individuals. In Fig.~\ref{fig3_prob} and \ref{fig4_evo2}, we examine the impact of bots on social cohesion across varying migration. Our findings indicate that random exploration behavior enhances bots' ability to promote cohesion within the population. Bots that remain static or lack random exploration (i.e., $p=0$) are unable to disrupt the frozen state (with $F_m$ remaining at 0), and fails to facilitate the segregation of defectors from cooperative clusters (with $\lambda_D$ remains around 0.45), thus demonstrating limited effectiveness in promoting cooperation, as depicted by the two leftmost columns in Fig.~\ref{fig4_evo2}. 
Conversely, when bots can randomly migrate exploration, even with a low level of exploration (i.e., $p=0.01$), they stimulate self-organizing movement and maintain a high level of aggregation among normal individuals, as shown in Fig.~\ref{fig3_prob}. Notably, bots with random migration (i.e., $p=1$)  can further elevate the migration level of normal individuals, resulting in a higher value of $F_m$ compared to $p=0.01$. This can drive the establishment of an exceptionally large cooperative cluster, as depicted in the rightmost column of Fig.~\ref{fig4_evo2}. However, under a high dilemma strength, bots with high-level exploration (i.e., $p > 0.7$) fail to promote cooperation. In contrast, bots with low-level exploration can still increases cooperation, as demonstrated in Fig.~\ref{fig3_prob} and Fig.~\ref{figs_r_rho} of SI.

Interactive action of bots is another critical factor influencing formation of social cohesion. When a population includes defective bots that consistently opt for unconditional defection, despite their potential to facilitate individual migration (as illustrated in Fig.~\ref{fig_evo_fcb} of SI where $F_m$ is non-zero), it does not lead to the separation of defectors from cooperative clusters. Moreover, the presence of defective bots diminishes the ability of cooperative bots to promote social cohesion. Enhanced social cohesion only occurs when the proportion of cooperative bots within the bot subgroup exceeds a certain threshold, while both defective and cooperative bots coexist, as depicted in Fig.~\ref{figs_fb} of SI.

\subsection{Robustness of model}

To evaluate the impact of action sequence in our model, we varied the sequence such that all individuals migrate before engaging in interactions and updating strategies \cite{vainstein2007does}. Results illustrate that altering the action sequence can enhance social cohesion when cooperative bots are involved, as shown in Fig.~\ref{figs_odr} of SI. However, under a high dilemma strength, bots without migration exploration outperform those with such exploration. When the action is limited, where individuals can only migrate or update strategies within a Monte Carlo time step, a minority of cooperative bots can still foster social cohesion, as depicted in Fig.~\ref{figs_mo} of SI. Moreover, we explore the influence of different lattice network topologies, where individuals have broader interaction and migration ranges, as well as larger population sizes. Results from Fig.~\ref{figs_size} and \ref{figs_deg} in the SI demonstrate the robustness of cooperative bots in promoting social cohesion across diverse lattice network topologies and population sizes.

\section{Conclusions and discussions}

This work employs evolutionary game theory to analyze how bot affect collective behavior of mobile population within networks. Results shown that cooperative bots can enhance the social cohesion among selfish, greedy individuals. When these individuals cannot find favorable migration positions, the lack of migration motivations leads to the emergence of frozen state in population \cite{helbing2009outbreak}. Interestingly, introduction of cooperative bots can break this frozen state. The random exploratory movements of bots create favorable positions, facilitating the clustering of cooperators and isolating defectors from cooperative clusters, thus leading to the defeat of defectors by tightly knit cooperative clusters. Even a minority of cooperative bots can promote social cohesion within the population, see Fig.~\ref{figs_phi} of SI. This suggests that cooperative bots act as internal forces social self-organization. The effect of cooperative bot is not limited by specific migration-imitation sequences or individual limited mobility, nor does it rely on the population size and the topology of normal network (refer to Fig.~\ref{figs_odr}, Fig.~\ref{figs_mo} and Fig.~\ref{figs_deg} in the SI). 

However, in extremely sparse populations, cooperative bots cannot help the establishment of cooperative clusters. In extremely densely populations, they fail to facilitate the separation of defectors from cooperative clusters during the self-organization process, thus diminishing their ability to eliminate defection. In our model, normal individual decisions rely on complete information regarding neighbors' behavior. When individuals lack information and resort to random movement, they rely on self-judgment to update strategies, cooperative bots still contribute to maintaining cooperation but cannot drive the aggregation of individuals. These conditions weaken the capacity of cooperative bots to promote social cohesion. Furthermore, our findings indicate that the presence of defective bots also hampers the establishment of social cohesion and diminishes the efficiency of cooperative bots in facilitating cooperation. These suggest the need for careful consideration in the design of bot behaviors. 

Our research is conducted within the context of one-shot games, where individuals make decisions without access to information regarding the past behavior of their co-players. While cooperative bots share some similarities with human zealots \cite{masuda2012evolution,cardillo2020critical,shen2023committed}—both consistently opting for cooperation—there are critical distinctions. Human zealots are rare in realistic settings, making it impractical to rely on them for widespread application if a high propensity for cooperation requires a substantial number of human zealots. In contrast, the behavior and scale of digital bots are controllable, making them effective tools for influencing human beliefs and behaviors in various online aspects, such as elections \cite{bessi2016social}, voting \cite{stella2018bots}, and political issues\cite{bail2018exposure}.

Our findings hold broad implications for online social platform, particularly concerning trust and opinion conflicts. For example, users often share opinions and collaborate with others online to accomplish tasks. However, misinformation and hostile communication environments frequently escalate conflicts, leading users to sever social connections and neglect collective interests. Using cooperative bots—designed to maintain friendly communication and provide collaborative assistance—can help create a more congenial communication environment and propagate collective consciousness \cite{traeger2020vulnerable}. Particularly, recent advancements in large language models (LLMs) exhibit impressive communication prowess and the potential to shape individuals' beliefs \cite{abdullah2022chatgpt}. This enables the construction of these cooperative agents to facilitate connectivity and communication among users, potentially enhancing cooperation and trust within online communities. Furthermore, our results highlight the critical role of incorporating random exploratory migration into bot design—allowing bots to roam different online communities—can help bridge connections among disconnected users and isolated communities to shape collective cohesion.

We employ a two-dimensional grid network, the simple network structure, though not a fully reflection of real-life social networks, encapsulates crucial social network features: limited interactions among individuals and engagement with neighbors. We anticipate that our findings remain robust, as they stem from the involvement of bots in individual limited interactions, which is independent of specific topological structures. Real-world network structures may display heterogeneity and time-varying \cite{strogatz2001exploring}, future investigations into these characteristics will enhance understanding of bot impacts.

In real-world, besides one-shot interactions, repeated interactions are also common. Bots with simple behaviors may not suffice to guide collective actions in this scenario. Instead, bots might be susceptible to manipulation and exploitation by humans, resulting in inefficiencies \cite{hilbe2014extortion}. Further consideration of memory-based strategy design may help explore the impact of bots on individuals in repeated games\cite{hilbe2014extortion,he2024impact,chen2023outlearning}. A key assumption in our study is that humans are unaware of interacting with bots, whether in game interactions or migration processes. When individuals become aware that their counterparts are bots, issues of trust in human-machine interactions \cite{ishowo2019behavioural} and biases towards bots \cite{karpus2021algorithm} emerge, which are critical factors affecting bot efficiency. Unfortunately, the impact of these factors remains unclear. Furthermore, we only focused on selfish and greedy individuals, as this aids in our exploration of whether bots alone can foster cooperation. However, human behavior is motivated by various factors beyond the pursuit of self-interest maximization. It is also influenced by social norms \cite{capraro2021mathematical} and various value-orientations \cite{schwarting2019social}. Future endeavors may benefit from integrating diverse behavioral decisions to comprehensively understand the impact of bots on collective behavior. Addressing these challenges will provide deeper insights to harness bots as effective tools in solving complex social issues.

\begin{acknowledgments}

\paragraph*{Funding} 
We acknowledge the support provided by (i) Major Program of National Fund of Philosophy and Social Science of China (Grants No.~22\&ZD158 and No.~22VRCO49) to L.S.; (ii) JSPS Postdoctoral Fellowship Program for Foreign Researchers (Grant No.~P21374), and an accompanying Grant-In-Aid for Scientific Research from JSPS KAKENHI (Grant No.~JP 22KF0303) to C.S.; (iii) the Grant-In-Aid for Scientific Research from JSPS, Japan, KAKENHI (Grants No.~JP 20H02314 and No.~JP 23H03499) awarded to J.\,T.; and (iv) China Scholarship Council (Grant No. ~202308530309) to Z.\,H.

\paragraph*{Author contributions} L.S., Z.H. and C.S. designed research; L.S., Z.H. and C.S. performed research; Z.H., C.S. and J.T. analyzed results; L.S., and J.T. supervision; L.S., Z.H., C.S. and J.T. wrote the paper. 

\paragraph*{Conflict of interest} Authors declare that no conflict of interests.

\paragraph*{Data accessibility} The code used in the study to produce results is freely available at  \href{https://osf.io/tu6cq}{https://osf.io/tu6cq}.

\end{acknowledgments}

\bibliographystyle{unsrt}
\bibliography{mainref}

\clearpage
\onecolumngrid

\section*{
Supplementary Information for \\
``Enhancing social cohesion with cooperative bots in societies of greedy, mobile individuals''}

\renewcommand\thesection{SI\arabic{section}}
\setcounter{section}{0}
\setcounter{figure}{0}
\renewcommand{\thefigure}{S\arabic{figure}}
\setcounter{page}{1}

This ``supporting information'' presents multiple variations of the basic model presented in the main text  to validate the effectiveness of simple bots in promoting social cohesion. Unless stated otherwise, we focus on scenarios representing low dilemma strength ($r=0.2$) and high dilemma strength ($r=0.4$), which respectively correspond to situations where cooperation can or cannot be sustained in a traditional mobile population without bots. We set the population density to $\rho=0.6$ and conducted the model on a lattice network with a size of $L=100$ and a Von Neumann neighborhood. In scenarios involving bots, the fraction of bots is set to $\phi=0.5$, and we examine four types of bots: static bots (i.e., unable to migrate), bots without random exploration (i.e., $p=0$), low-exploration bots (i.e., $p=0.01$), and bots with random movement (i.e., $p=1$). The following sections are organized as follows: Sec. \ref{si_eviormen} presents the results of bots under different population densities $\rho$ and dilemma strengths $r$. Sec. \ref{si_bot} reports the results of various bot settings in terms of strategy and fraction of bots. Sec. \ref{si_inf} reports he influence of information level in individual decision making. Sec. \ref{si_beh} discusses the effect of bots in the presence of individual behavioral noise. Sec. \ref{si_odr} covers the results when there are changes in migration and strategy update sequences. Sec. \ref{si_size} examines the effect of network size and topology. The robustness of our conclusions is demonstrated through various results based on these variant models. The code used in the study to produce computer simulation results is freely available at  \href{https://osf.io/tu6cq}{https://osf.io/tu6cq}.

\section{The effect of cooperative bots on social cohesion  across varying population densities\label{si_eviormen}}

\begin{figure}[htbp]
    \centering \includegraphics[width=0.8\linewidth]{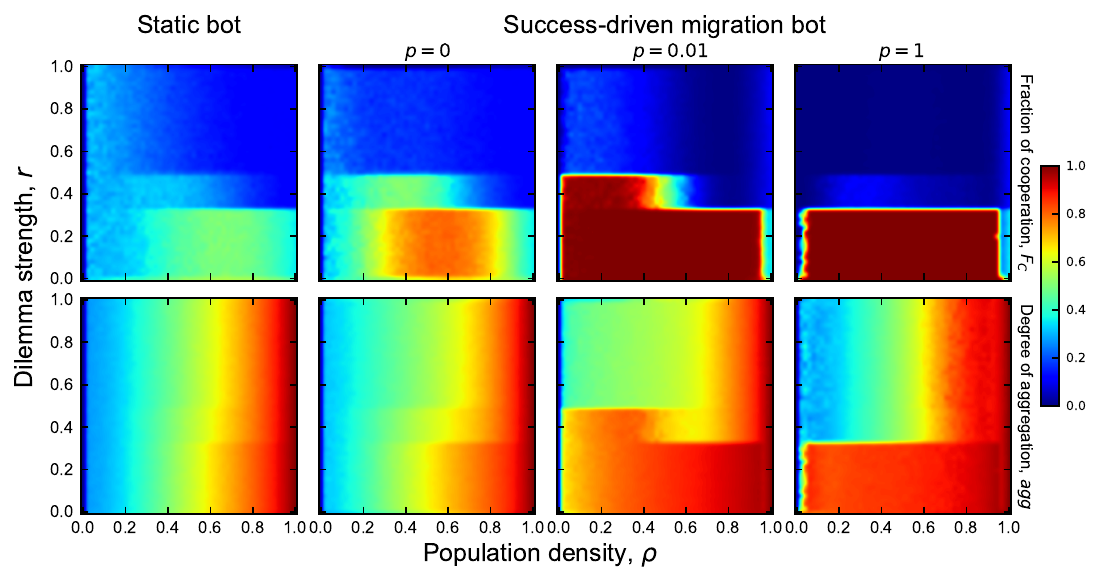}
    \caption{ The color code indicates the fraction of cooperation (top panels) and the degree of aggregation of normal individuals (bottom panels) as a function of population density $\rho$ and dilemma strength $r$. }
    \label{figs_r_rho}
\end{figure}

Fig. \ref{figs_r_rho} illustrates the impact of four representative types of bots on social cohesion. As expected, there exists an optimal population density for each bot to enhance cooperation. Bots with exploratory migration prove most effective in promoting social cohesion at low dilemma strengths ($r < 1/3$), resulting in pure cooperation over a wide parameter range of $\rho$. Low-exploration bots and success-driven bots perform well at high levels of dilemma strength ($1/3 < r < 1/2$) since their random movements can stimulate individual migration. However, completely random movements of bots harm close cooperative clusters, particularly at high temptation levels, while low-exploration is beneficial to the aggregation of cooperative individuals.

\begin{figure}[htbp]
    \centering
    \includegraphics[width=0.8\linewidth]{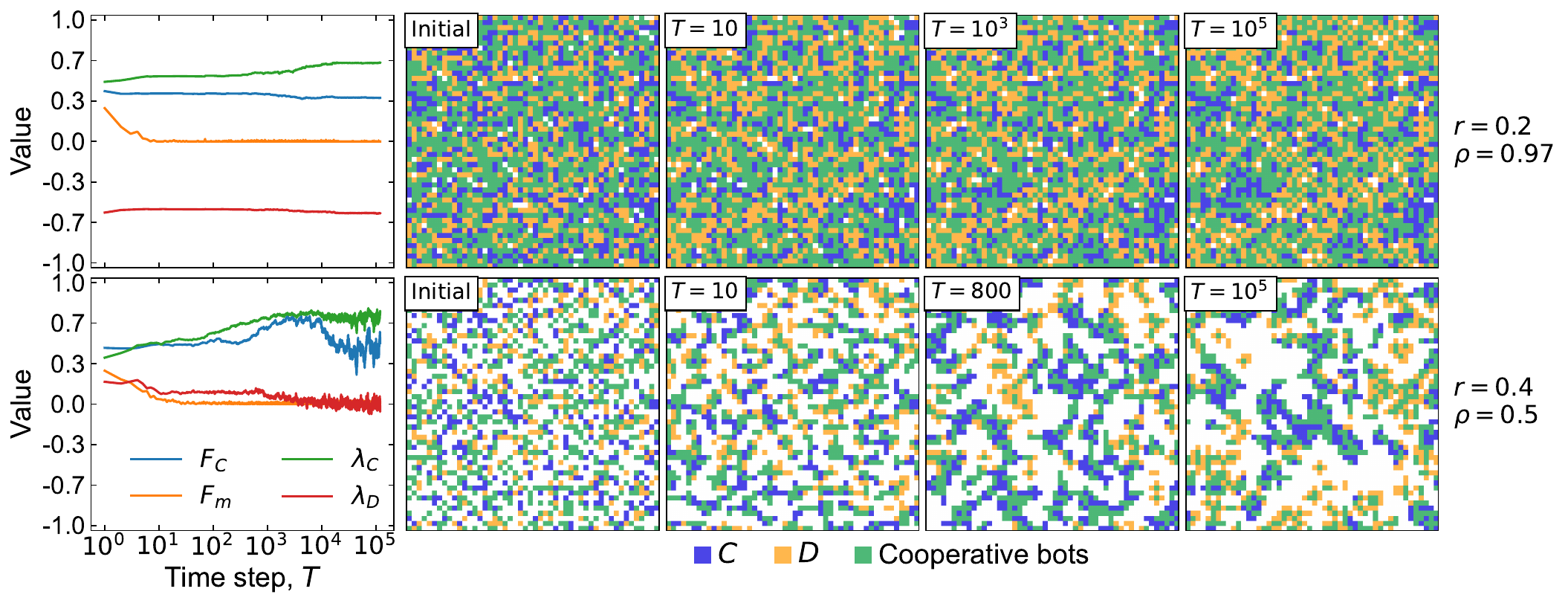}
    \caption{A high population density or a high level of dilemma strength can diminish the effectiveness of cooperative bots in completely eliminating defection within self-organizing movements. Leftmost panels show the fraction of cooperation $F_C$, the fraction of normal individuals who have moved $F_m$, clustering level of normal cooperators $\lambda_C$ and the isolation level of normal defectors $\lambda_D$ as a function of time steps. The right panels showcase evolutionary snapshots in scenarios with and without bot, respectively. In these snapshots, blue, yellow, green, and white dots represent cooperators, defectors, bots, and empty sites, respectively. Results are obtained by setting $p=0.01$.}
    \label{fig_evo_condi}
\end{figure}

Our research demonstrates that cooperative bots effectively can hinder defection and foster pure cooperation under population with a moderate density. However, this effectiveness wanes in high-density populations or in scenarios with high dilemma strengths. In extremely densely populated settings, as depicted in the top panels of Fig.~\ref{fig_evo_condi}, snapshots reveal defectors closely intertwined with cooperators ($\lambda_D < 0 $ is evident), indicating the inability of cooperative bots to segregate defectors from cooperative clusters through self-organization. Consequently, defection persists even under low dilemma strength. In scenarios with high dilemma strength, illustrated in the bottom panels of Fig.~\ref{fig_evo_condi}, although bots facilitate the segregation between defectors and cooperators ($\lambda_D > 0 $ is evident), defection persists due to the it's advantageous payoff, rendering complete eradication unachievable.

\section{The influence of bot sub-population size and interactive actions on social cohesion \label{si_bot}}

\begin{figure}[htbp]
    \centering
    \includegraphics[width=0.8\linewidth]{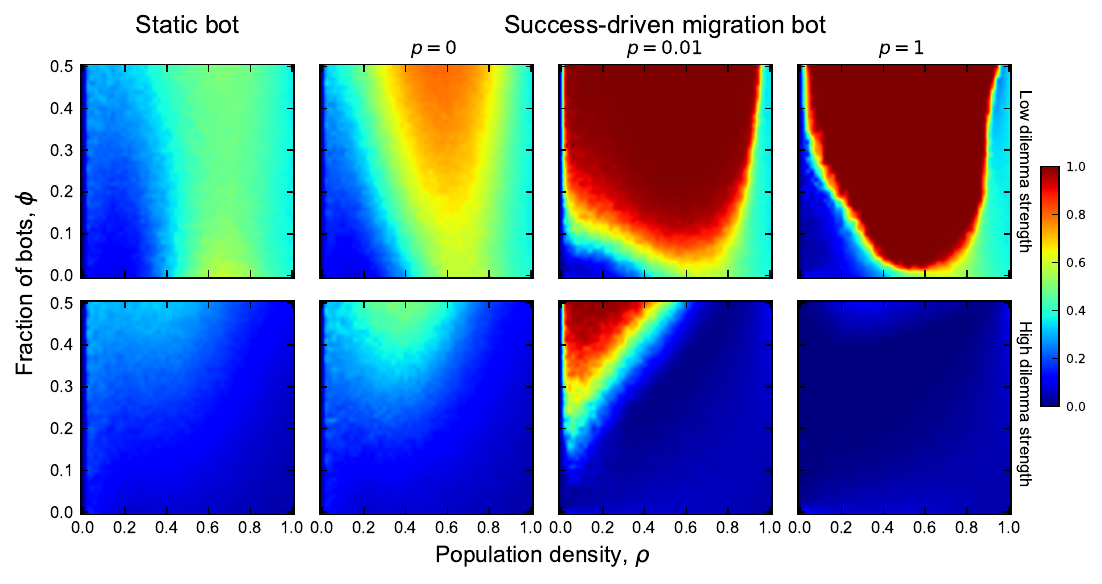}
    \caption{The color code indicates the degree of aggregation of normal individuals as a function of population density $\rho$ and the fraction of bots $\phi$.}
    \label{figs_phi}
\end{figure}

\begin{figure}[!ht]
    \centering
    \includegraphics[width=0.8\linewidth]{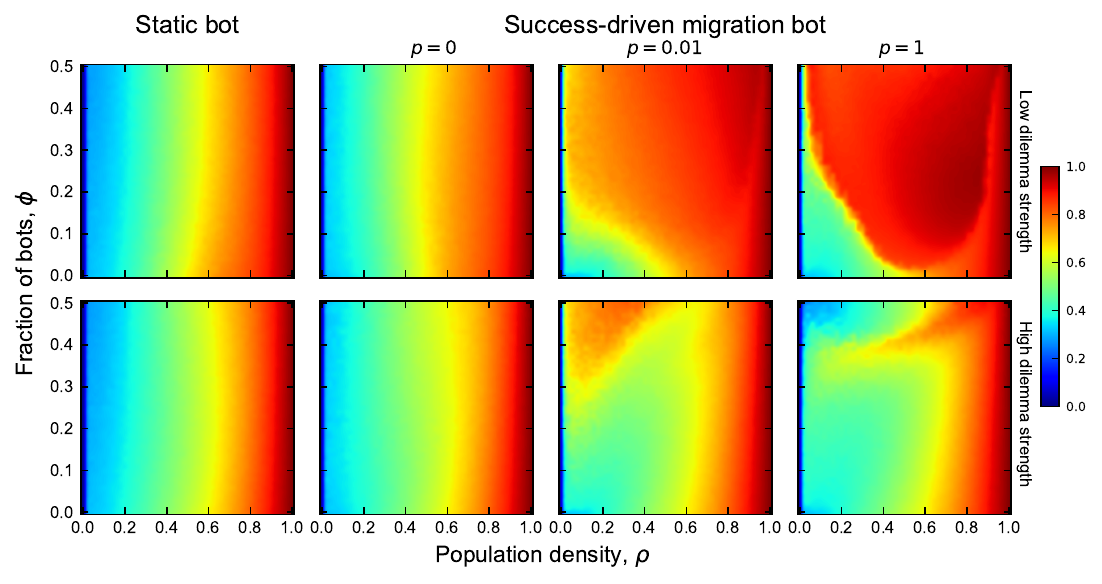}
    \caption{The color code indicates the degree of aggregation of normal individuals as a function of population density $\rho$ and the fraction of bots $\phi$.}
    \label{figs_phi_agg}
\end{figure}

The results from Fig.~\ref{figs_phi} and \ref{figs_phi_agg} demonstrate that a higher presence of cooperative bots can promote social cohesion. Interestingly, at moderate population densities ($\rho=0.6$), a mere proportion of $\phi=0.07$ cooperative bots is adequate to eradicate defection and induce a state of pure cooperation, whereas the proportion of low-exploration bots needs to reach $0.16$. In scenarios with high dilemma strength, a minimum proportion of 0.2 low-exploration bots is necessary to facilitate heightened social cohesion. Across all population densities, elevating the ratio of bots proves advantageous in enhancing cooperative behavior.

\begin{figure}[htbp]
    \centering
    \includegraphics[width=0.6\linewidth]{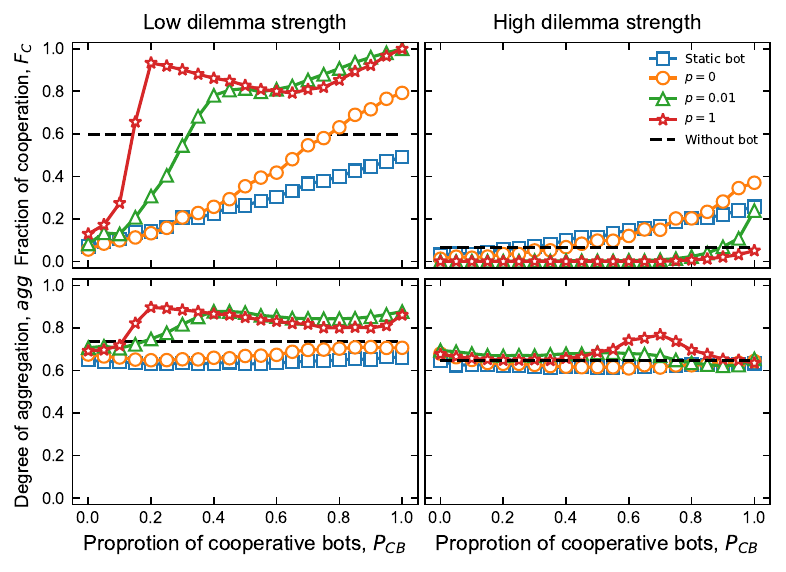}
    \caption{The fraction of cooperation $F_C$ (top panels) and the degree of aggregation $Agg$ (bottom panels) as a function of the fraction of cooperative bots $F_{BC}$. Considering the introduction of defective bots which consistently adopt unconditional defection in game interactions, within a bot sub-population, where the proportion of cooperative bots is represented by $P_{CB}$ and the proportion of defective bots is $1-P_{CB}$. }
    \label{figs_fb}
\end{figure}

\begin{figure}[htbp]
    \centering
    \includegraphics[width=0.8\linewidth]{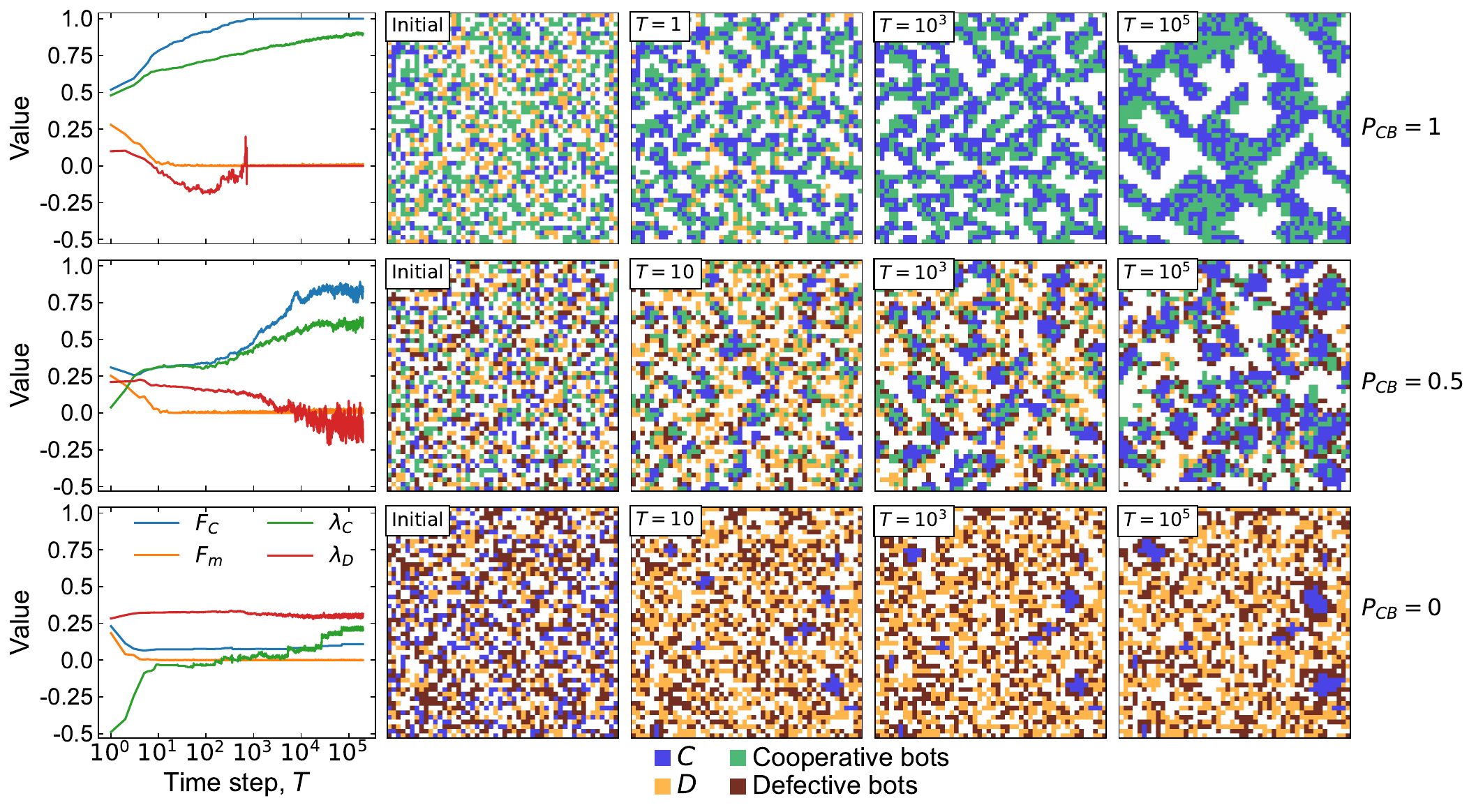}
    \caption{
    Leftmost panels show the fraction of cooperation $F_C$, the fraction of normal individuals who have moved $F_m$, clustering level of normal cooperators $\lambda_C$ and the isolation level of normal defectors $\lambda_D$ as a function of time steps. The right panels showcase evolutionary snapshots in scenarios with and without bot, respectively. In these snapshots, blue, yellow, green, brown and white dots represent cooperators, defectors, cooperative bots, defective bots and empty sites, respectively. In the bot sub-population, the proportion of cooperative bots is $p_{CB}$, while the proportion of defective bots which consistently choose unconditional defection is $1-p_{CB}$. Results are obtained by setting $\rho = 0.6$, $r=0.2$ and $p=0.01$.}
    \label{fig_evo_fcb}
\end{figure}

To explore the influence of bots’ action in game interactions, we introduce defective bots within bot sub-population, consistently employing unconditional defection strategies. The outcomes, illustrated in Fig.~\ref{figs_fb}, demonstrate a decline in social cohesion attributed to these defective bots. The presence of defective bots indeed facilitates individual movement, resulting in non-zero migration proportions (i.e., $F_m>0$). Although this also can disrupt the frozen state of population, defective bots do not contribute to enhancing social cohesion. In fact, these defective bots impede the separation of normal defectors and cooperators during the self-organization process (as indicated by negative values of $\lambda_D$ in the middle panels of Fig.~\ref{fig_evo_fcb}) and hinder the expansion of cooperative clusters (as shown in the bottom panel of Fig.~\ref{fig_evo_fcb}). This hindrance not only fails to improve social cohesion but also diminishes the ability of cooperative bots to enhance social cohesion.

\section{The impact of limited complete information\label{si_inf}}

\begin{figure}[htbp]
    \centering
    \includegraphics[width=0.6\linewidth]{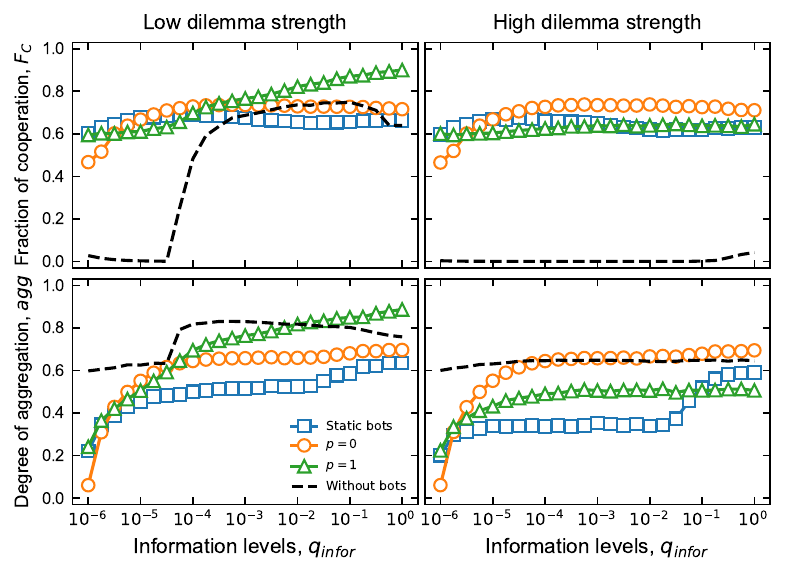}
    \caption{The fraction of cooperation $F_C$ (top panels) and the degree of aggregation $Agg$ (bottom panels) as a function of the information level $q_{infor}$ under a moderate population with $\rho=0.6$. Under constraints of acquiring complete information, normal players may have a probability $q_{infor}$ of obtaining complete information when updating their strategies. In these cases, normal individuals can employ the ``best-take-over'' rule to mimic the most successful strategy, otherwise, they rely on the myopic principle to update their strategies. Similarly, during migration, normal players may potentially follow a ``success-driven'' rule with a probability $q_{infor}$ of acquiring complete information; otherwise, they resort to random movement due to the absence of information.}
    \label{figs_infor}
\end{figure}

In our model, the decision-making of normal individuals, whether in strategy updating or migration, depends on having complete information about the strategies of surrounding players. To explore the effects of limited information acquisition, we introduce the parameter $q_{infor}$  to represent the likelihood of complete information being available. Specifically, normal players have a probability $q_{infor}$ of acquiring complete information when updating their strategies, allowing them to adopt the ``best-take-over'' rule to mimic the most successful strategy. Otherwise, they rely solely on their own judgment. In the absence of complete information, they switch strategies with a probability $w_{s\to s^{'}}$ following a myopic principle, as given by:
\begin{equation}
    w_{s\to s^{'}} = \frac{1}{ 1 + \exp \{  \beta ( R_{s} - R_{s^{'}}  ) \} },
\end{equation}
where $R_{s}$ represents the player's current payoff from adopting strategy $s$, $R_{s^{'}}$ is the expected payoff when another strategy $s^{'}$ is adopted, and $\beta$ is a sensitivity parameter (we set $\beta = 10$ to indicate strong dependence on payoff differentials). Similarly, during migration, normal players may follow the "success-driven" rule with a probability $q_{infor}$ of obtaining complete information; otherwise, they engage in random movement due to insufficient information. Fig.~\ref{figs_infor} demonstrates that mobile cooperative bots, promoting cooperation, remain effective regardless of the limited information. However, the decrease in information levels prevents cooperative bots from promoting individual aggregation.

\section{
The impact of normal individual behavior noise \label{si_beh} }

\begin{figure}[htbp]
    \centering
    \includegraphics[width=0.6\linewidth]{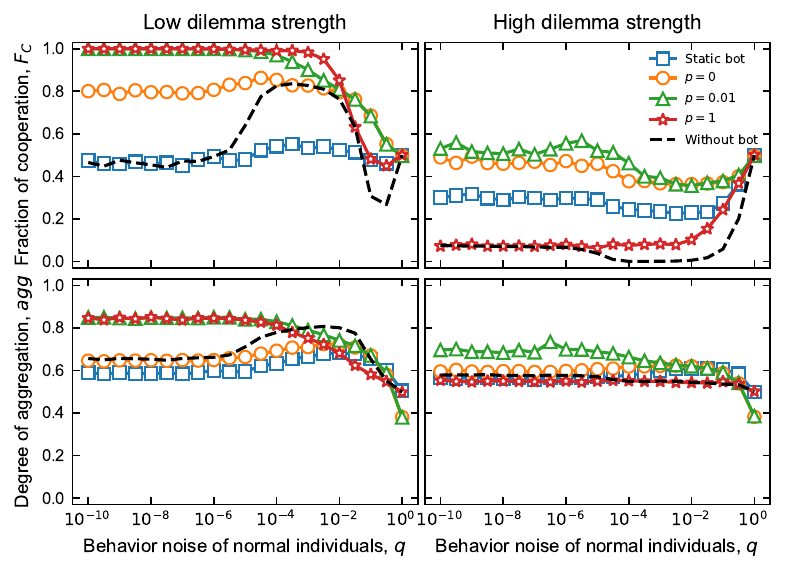}
    \caption{The fraction of cooperation $F_C$ (top panels) and the degree of aggregation $Agg$ (bottom panels) as a function of the behavioral noise of normal individuals $q$ under a moderate population with $\rho=0.6$. Considering the presence of behavioral noise in individual decision-making, this implies that individuals have a probability $q$ of randomly resetting their strategies when updating interaction strategies, and similarly, a probability $q$ of randomly selecting migration destinations during migration.
    }
    \label{figs_nois}
\end{figure}

\begin{figure}[htbp]
    \centering
    \includegraphics[width=0.6\linewidth]{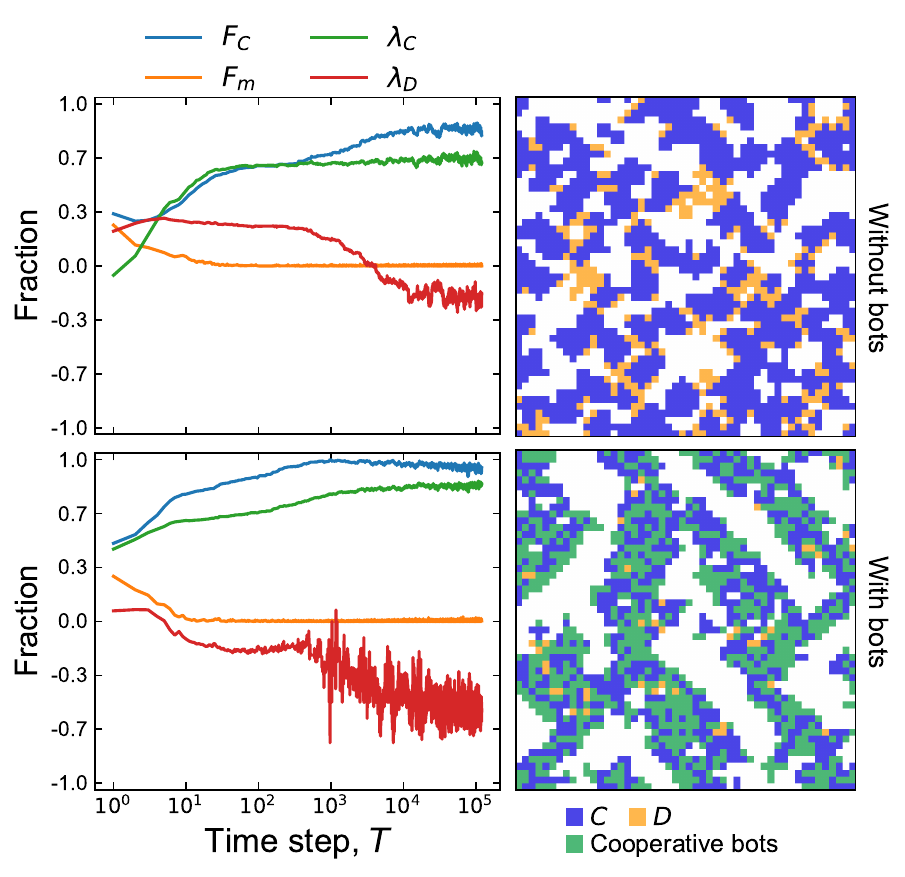}
    \caption{Left panels show the fraction of cooperation $F_C$, the fraction of normal individuals who have moved $F_m$, clustering level of normal cooperators $\lambda_C$ and the isolation level of normal defectors $\lambda_D$ as a function of time steps. Right panels show that stable spatial distribution in scenarios with and without bot, respectively. Blue, yellow, green, and white dots represent cooperators, defectors, bots, and empty sites, respectively. Results are obtained by setting $\rho = 0.5$, $r=0.2$, $p=0.01$ and $q=10^{-4}$.}
    \label{figs_evo}
\end{figure}

The findings of this study are based on the assumption that individuals make rational decisions. We also consider relaxing this assumption by taking into account behavioral noise in decision-making. This includes individuals randomly resetting their strategy with probability $q$ during strategic updates and carrying out random movement in the migration stage with probability $q$, as shown in Fig. \ref{figs_nois}. Our findings reveal the significant role of cooperative bots even in the presence of behavioral noise, as they contribute to stabilizing cooperative clusters (see bottom panels of Fig. \ref{figs_evo}). Dynamic demonstrations for a visual understanding of the evolutionary process are available online at: \hyperlink{https://osf.io/tu6cq}{https://osf.io/tu6cq}. In traditional cases, moderate behavioral noise proves helpful in breaking individuals out of the frozen state of migration (see top panels of Fig. \ref{figs_evo}). Interestingly, the presence of cooperative bots still serves to stabilize cooperative clusters, thus slowing down the decline of cooperation. When compared to scenarios without bots, mobile bots can maintain a higher fraction of cooperation and degree of aggregation.

\section{
The impact of changing the order of action and limited actions\label{si_odr}}

\begin{figure}[htbp]
    \centering
    \includegraphics[width=0.6\linewidth]{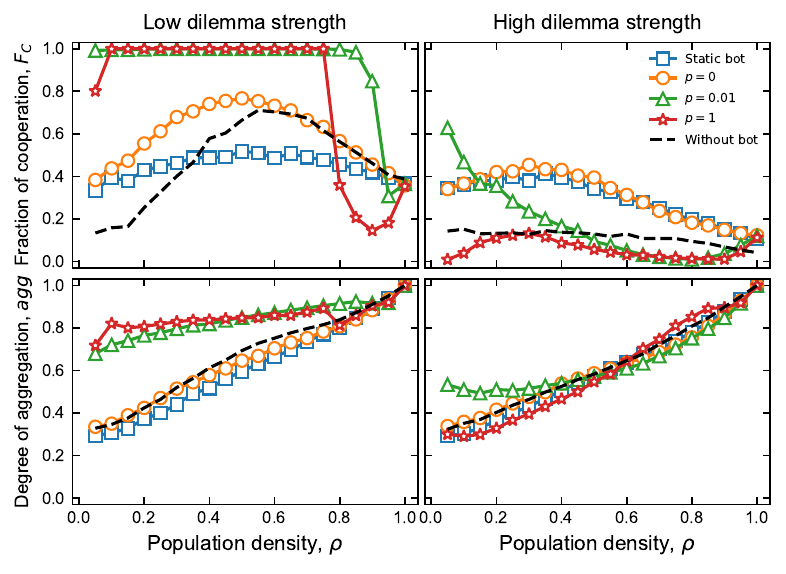}
    \caption{The fraction of cooperation $F_C$ (top panels) and the degree of aggregation $Agg$ (bottom panels) as a function of population density $\rho$, with changes in the order of strategic updating and migration stages. In this scenario, all individuals migrate first, followed by interactions to gain profits and update strategies.}
    \label{figs_odr}
\end{figure}

\begin{figure}[htbp]
    \centering
    \includegraphics[width=0.8\linewidth]{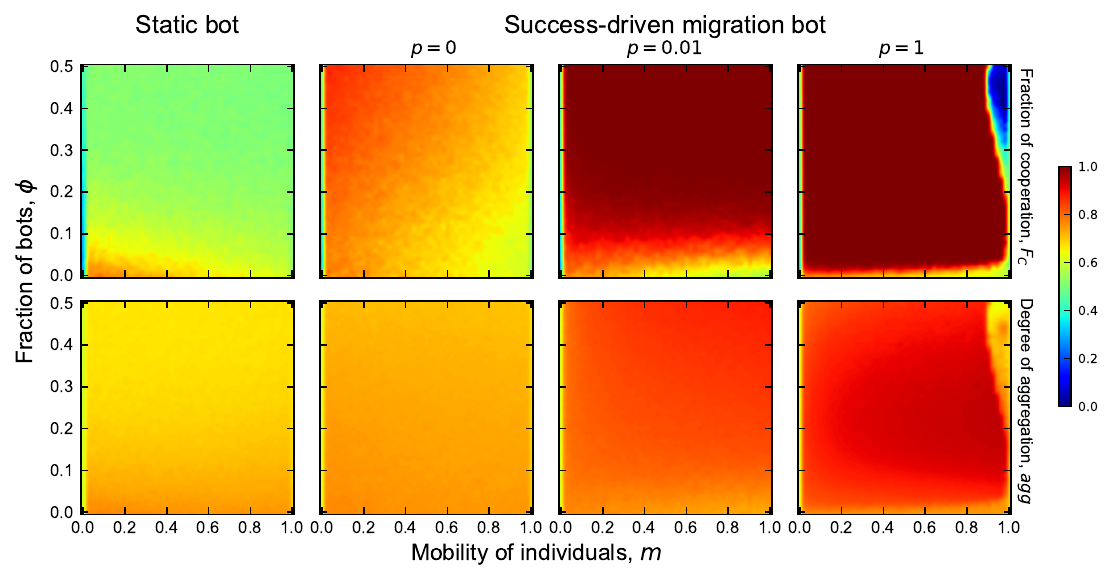}
    \caption{The color code indicates the $F_C$ (top panels) and the degree of aggregation $Agg$ (bottom panels) as a function of the fraction of bots $\phi$ and individuals' mobility $m$. This scenario accounts for the constrained mobility of individuals, where at each Monte Carlo time step, an individual can only choose between migration or strategy imitation. Hence, individual mobility is defined as the probability $m$ of making migration decisions per round, or the probability $1-m$ of updating strategies. Parameters are set to $\phi=0.5$ and $r=0.2$. }
    \label{figs_mo}
\end{figure}

In Fig. \ref{figs_odr}, we present the results of exchanging the sequence of strategic updates and migration stages. In this scenario, normal individuals migrate first and then interact with their neighbors to update their strategies at each MC time step. Cooperative bots can still promote cooperation even when the order of strategic updating and migration is changed. Low-exploration bots prove to be particularly robust in facilitating cooperation. Additionally, we investigate a model where normal individuals either migrate with a probability $m$ or perform strategy updates with a probability $1-m$, resulting in inconsistent time scales for individual strategy updating and migration. The consistent findings are depicted in Fig. \ref{figs_mo}.

\section{The impact of network size and topology \label{si_size}}

\begin{figure}[htbp]
    \centering
    \includegraphics[width=0.8\linewidth]{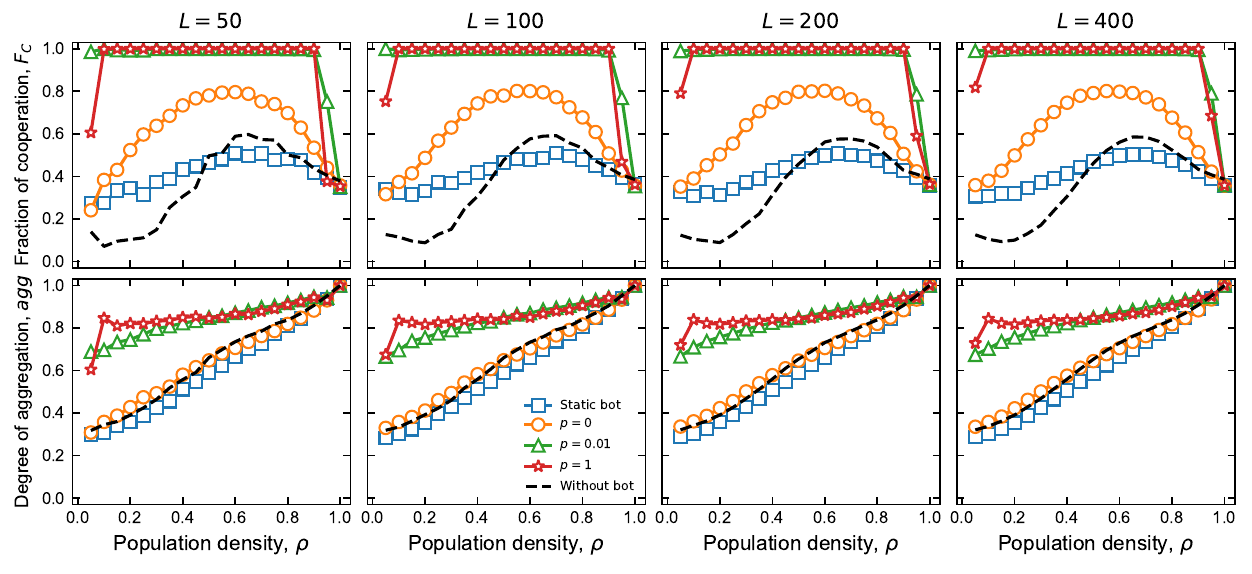}
    \caption{The fraction of cooperation $F_C$ (top panels) and the degree of aggregation $Agg$ (bottom panels) as a function of population density under various sizes of lattice networks with Von Neumann neighborhoods. Parameter is set to $r=0.2$.}
    \label{figs_size}
\end{figure}

\begin{figure}[htbp]
    \centering
    \includegraphics[width=0.8\linewidth]{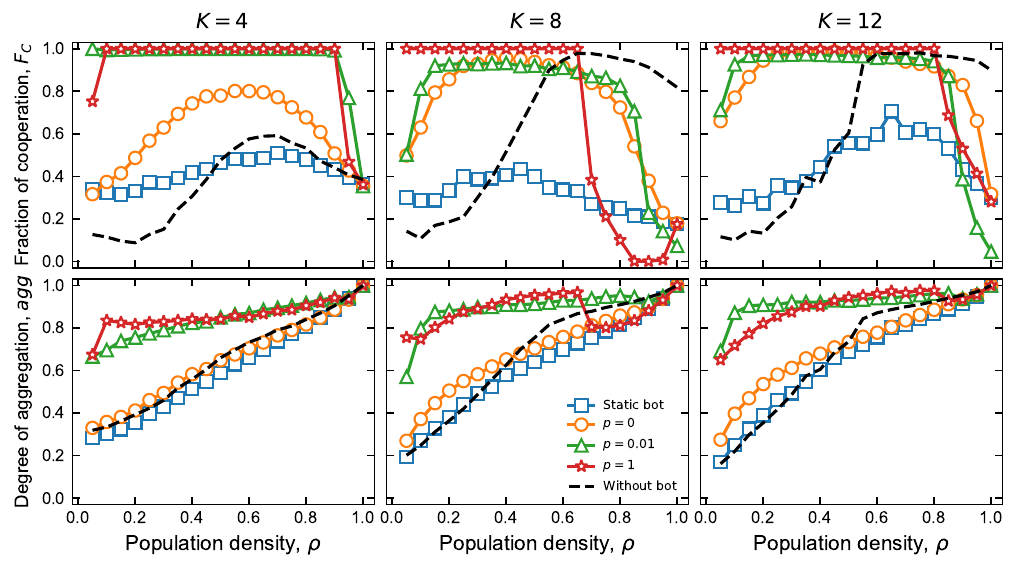}
    \caption{The fraction of cooperation $F_C$ (top panels) and the degree of aggregation $Agg$ (bottom panels) as a function of the population density under grid lattice network with $K=4$ (Von Neumann neighborhood), $K=8$ (Moore neighborhood) and $K=12$. In these networks, individuals engage in game interactions and migrate within immediate K-nearest locations. Parameter is set to $r=0.2$.}
    \label{figs_deg}
\end{figure}

Fig.~\ref{figs_size} depicts consistent outcomes across diverse network sizes in grid networks with Von Neumann neighborhoods, underscoring the resilience of bot influence concerning population size. Expanding our analysis to lattice networks with $k=8$ (i.e., Moore neighbors) and $k=12$ broadens individual interaction and migration decision scopes. Despite modifications in lattice network structures, bots still exhibit their capacity to promote social cohesion. However, in a densely populations, bots' ability to foster social cohesion diminishes, instead hindering it. With expanded interaction and migration ranges, the segregating impact of network structures weakens, thereby diminishing bots' effectiveness in distinguishing defection from cooperators. Under conditions of low population density, cooperative bots can still significantly promote social cohesion.

\clearpage
\section*{Supplementary figures}


\end{document}